\documentclass[aps,pra,reprint,superscriptaddress,floatfix]{revtex4-1}
\usepackage{graphicx,amsmath,amssymb}
\usepackage{graphics}
\usepackage{bm}
\usepackage{color}

\newcommand{\ket}[1]{\lvert{#1}\rangle}

\begin{document}


\title{Loading of a fountain clock with an enhanced Low-Velocity Intense Source of atoms}


\author{G.~Dobrev}
\affiliation{Physikalisch-Technische Bundesanstalt, Bundesallee 100, 38116 Braunschweig, Germany}
\affiliation{Faculty of Physics, Sofia University, 5 J.Bourchier Blvd, 1164 Sofia, Bulgaria}
\affiliation{ISSP, Bulgarian Academy of Sciences, 72 Tzarigradsko Chaussee Blvd, 1784 Sofia, Bulgaria}
\author{V.~Gerginov}
\affiliation{Physikalisch-Technische Bundesanstalt, Bundesallee 100, 38116 Braunschweig, Germany}
\author{S.~Weyers}
\affiliation{Physikalisch-Technische Bundesanstalt, Bundesallee 100, 38116 Braunschweig, Germany}


\date{\today}

\begin{abstract}
We present experimental work for improved atom loading in the optical molasses of a caesium fountain clock, employing a low-velocity intense source of atoms (LVIS) [Lu \emph{et al.}, Phys.~Rev.~Lett.~\textbf{77},~3331~(1996)], which we modified by adding a ``dark'' state pump laser. With this modification the atom source has a mean flux of $4 \times 10^{8}$ atoms/s at a mean atom velocity of $8.6$\,m/s. Compared to fountain operation using background gas loading, we achieved a  significant increase of the loaded and detected atom number by a factor of 40. Operating the fountain clock with a total number of detected atoms $N_{\mathrm{at}}=2.9 \times 10^6$ in the quantum projection noise-limited regime, a frequency instability $\sigma_y\left(1\text{s}\right)=2.7 \times 10^{-14}$ was demonstrated. \end{abstract}


\maketitle

\section{\label{sec:Intro}Introduction}

The total measurement uncertainty of frequency measurements performed with fountain primary frequency standards is obtained from the quadratic sum of the statistical and the  systematic measurement uncertainties \cite{Wynands2005}. A higher fountain frequency stability results directly in an improved statistical uncertainty in a given measurement time, accompanied by an improved  total measurement uncertainty. Moreover, there are systematic uncertainty contributions such as those from the collisional or the distributed cavity phase shifts, which can also be reduced with an improved statistical uncertainty of their evaluation \cite{Gerginov2010, Weyers2012b}. Thus an improved fountain stability during a given measurement can also lower the total measurement uncertainty by indirectly reducing the systematic uncertainty.

\begin{figure}
    \includegraphics[width=0.80 \columnwidth]{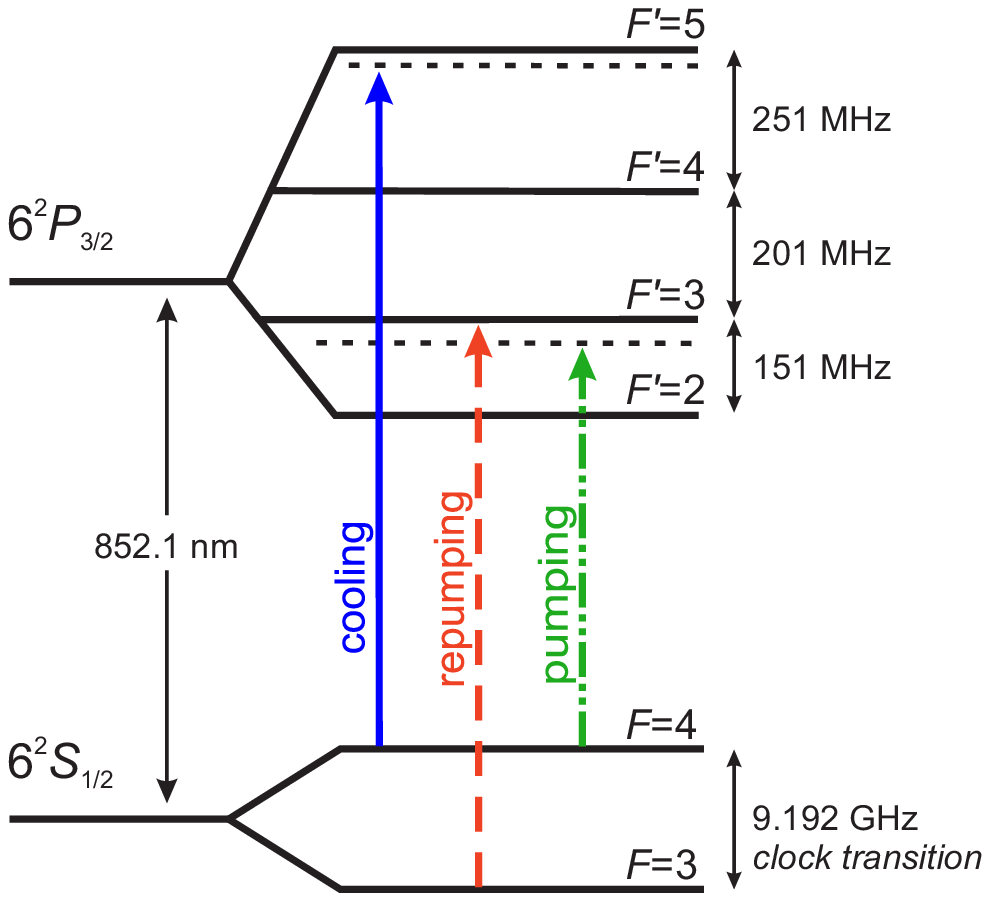}
    \caption{\label{Cs}The caesium D$_{2}$ line and the frequency splitting between the hyperfine components of the $6^{2}S_{1/2}$ ground and $6^{2}P_{3/2}$ excited states (not to scale). The transitions driven by the various laser fields, present in our LVIS trap, are shown as well.
}
\end{figure}

We describe improvements of the frequency stability of the fountain primary frequency standard CSF2 \cite{Gerginov2010}. In CSF2 caesium atoms are cooled and accumulated in an optical molasses (OM). The captured atoms are subsequently launched in vertical direction to perform frequency measurements of the microwave clock transition between the $6^{2}S_{1/2}$ ground state hyperfine sublevels $\ket{F=3}$ and $\ket{F=4}$ (Fig.~\ref{Cs}) \cite{Wynands2005}. With the use of an optically-stabilized microwave signal \cite{Weyers2009} the stability of the fountain is limited by quantum projection noise (QPN) over a wide range of atom numbers. As a result, the fountain frequency stability is improved with the square root of the detected atom number. The loaded and detected cold atom number is increased (and the corresponding QPN is reduced) by OM loading from a low-velocity intense source of cold atoms (LVIS, \cite{Wieman96}). The LVIS system is modified similar to the work of Teo \textit{et al.} \cite{teo2002} by adding a ``dark'' state pump laser. In this work the pump laser is used to reduce the velocity of the slow beam, and gives another factor of two loading enhancement.

The LVIS arrangement \cite{Wieman96} is very similar to the classical $3$D magneto-optical trap (MOT) scheme \cite{Metcalf} for atom cooling with an additional leak channel for the cold atoms.
In the LVIS scheme one of the six cooling laser beams has significantly reduced field intensity in a small central region of its spatial profile, because it is reflected from a mirror with a small central hole acting as atomic beam aperture. This feature perturbs the MOT trapping potential and creates the leak channel. Atoms in this region become subject to acceleration by the laser beam (in the following called ``accelerating beam'') pointing in the direction of the aperture. As a result, they get pushed out of the trap and form a cold continuous atom beam.

After a description of our experimental setup and the initial optimization procedures, we will present and explain the findings obtained from the insertion of a ``dark'' state pump laser. The cold atom beam from the enhanced LVIS system is characterized regarding beam flux and mean atom velocity, before we demonstrate the improved CSF2 frequency stability by evaluating frequency measurement results.

\section{\label{sec:Experiment}Experiment}

\subsection{\label{sec:Setup}Experimental setup}



\begin{figure}
    \includegraphics[width=0.48\textwidth]{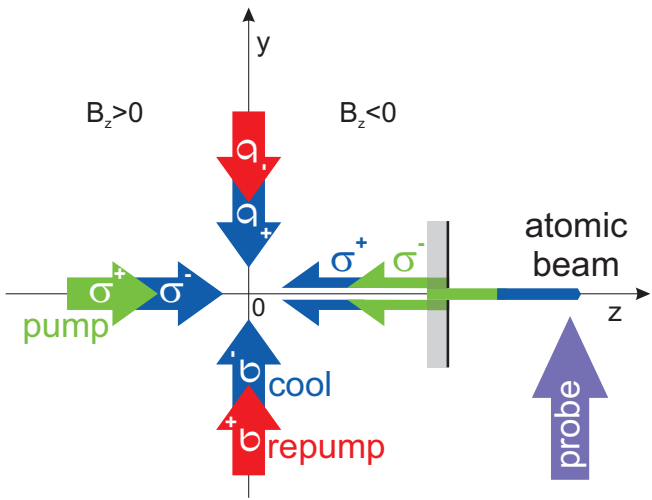}
    \caption{\label{LVIS}Schematic drawing of the enhanced low velocity intense source system. The magnetic field gradient coils are not shown.}
\end{figure}

The LVIS trapping zone is constructed around a standard DN35CF six--way cube. Optical access is provided by five AR-coated windows with diameter of $38$\,mm. The free end of the cube is attached to the OM chamber of the fountain vacuum system through a six-way cross and a flexible metal bellow. The distance between both trap centers is approximately $53$\,cm and the LVIS cube is positioned $11$\,mm higher than the OM center in order to compensate the height difference acquired by the atoms along their ballistic flight. Two identical coaxial coils, arranged in anti-Helmholtz configuration create the quadrupole magnetic field for the trap operation. The line passing through the coil centers defines the trap symmetry axis $z$ (see Fig.~\ref{LVIS}).

Diode lasers provide the trapping light. It is introduced into the LVIS trough five individual single-mode polarization maintaining fibers with collimators (Sch\"{a}fter\&Kirchhoff Fiber Collimator 60FC-T-4-M100-37) at the fiber ends. The collimators come with integrated quarter-wave plate and provide circularly polarized, collimated laser light having a gaussian profile with $21$\,mm beam diameter ($1/e^2$). Along the z axis, there is only one fiber collimator mounted, while on the opposite side of the MOT center, at a distance of $18$\,mm, the output coupler of the LVIS system is positioned. It consists of a pierced quarter-wave plate, with an $0.5$\,mm diameter aperture in the center and with a high-reflection coating deposited on the back side. The output coupler produces a retro-reflected laser beam which has opposite circular polarization with respect to the incident beam, which is a necessary condition for the MOT operation. The aperture in the output coupler forms the desired extraction column in the LVIS. No additional collimation of the atomic beam is performed.

\subsection{\label{sec:Opt}Initial optimization}

The number of cold atoms loaded in the OM zone of the fountain depends on the atomic beam flux, beam divergence and losses during the atom capture process in the OM. The atomic beam flux is determined mainly by the LVIS trap capture rate \cite{Wieman96}. Parameters relevant to the capture rate are the value of the atomic vapour background pressure, magnetic field gradient, cooling laser power, frequency detuning, intensity profile and distribution among the cooling laser beams, as well as repump laser parameters. An additional factor is the size of the aperture in the LVIS output coupler. The complexity of the system LVIS -- OM does not allow to fully separate the individual impact of all these factors on the loading process. We assess and optimize the performance of the overall system by observing the total number $N_{\mathrm{at}}$ of detected cold atoms at the end of the fountain interrogation cycle.

\begin{figure}

    \includegraphics[width=0.40\columnwidth]{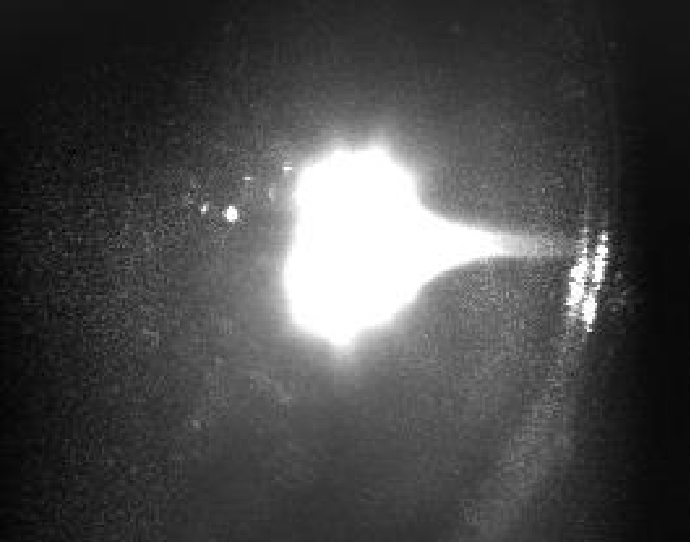}
    \caption{\label{cloud}A snapshot of the LVIS MOT region made with a CCD camera. The bright central spot is the fluorescence from the trapped caesium atoms, and the tail emerging from it is due to scattered photons from atoms which leave the trap and form the cold atom beam. The conditions at which the picture is taken are $18$\,mW cooling laser optical power on each beam, 0.79\,mT/cm axial and 0.39\,mT/cm radial magnetic field gradients.}

\end{figure}

Initially, the formation of an ensemble of cold atoms in the LVIS MOT was accomplished by distributing the power of the cooling laser equally among the three trap axes. The magnetic field orientation of the MOT defines the needed polarization state of each individual beam. In Fig.~\ref{cloud} one can see the fluorescence emitted by the trapped atoms in the LVIS MOT and the asymmetric shape of the cold atom cloud as a result of the imbalanced radiation pressure force along the $z$ axis. Fluorescence from atoms which form the atom beam and leave the MOT central region appears as a tail on the right hand side of the main cooled ensemble. Already without any further optimizations we encountered a noticeable change in $N_{\mathrm{at}}$ which indicated an enhanced loading of atoms in the OM zone of the fountain. An advantage of using the above described source of cold atoms is that the small aperture size in the LVIS output coupler provides a differential pumping mechanism between both trap chambers. This allowed us to increase the caesium partial pressure in the LVIS MOT zone and at the same time to preserve the rest of the fountain vacuum system from unfavourable rise of the local background pressure. This in turn would increase the OM loss rate, and the rate of  collisions of the interrogated cold atoms with the hot background Cs atoms during the free propagation time.

A Fabry-Perot laser diode (JDSU 5430) at $852$\,nm serves as LVIS cooling laser. Its frequency is stabilized through an injection locking technique to the frequency of the master laser of CSF2. An acousto-optic modulator is used to tune the laser absolute frequency about $2 \Gamma$ to the red of the $\ket{F=4}$ $\rightarrow$ $\ket{F'=5}$ cycling transition ($\Gamma=5.2$\,MHz is the natural linewidth of the transition). In total, the cooling laser delivers $90$\,mW optical power to the LVIS trap. The trap center is aligned with the extraction column by suitably distributing the laser power between the transverse laser beams.



The intensity of the accelerating beam is a crucial parameter, concerning the efficiency of the OM loading process. In our experimental arrangement the optimum value for the power of the accelerating beam was found to be $10$\,mW. Also the optimum value of the quadrupole magnetic field gradient is close to the one for optimum MOT capture rate. Beside laser parameters the value of the magnetic field gradient defines the rate of absorption-emission cycles that an accelerated atom experiences and thus the final atom beam velocity. In our experiment, the anti-Helmholtz coils are driven with a current of $2.7$\,A, creating $0.71$\,mT/cm and $0.35$\,mT/cm magnetic field gradients in axial and radial directions, respectively. An additional distributed Bragg reflector (DBR) laser diode (Photodigm PH852DBR120) is used as a repump laser to bring the atoms which have decayed to the $\ket{F=3}$ state back to the cooling cycle. The DBR laser diode delivers a total optical power of $6.2$\,mW into the LVIS MOT and its output frequency is stabilized by saturated absorption spectroscopy. This laser shares the same optical fibers with the cooling laser depending on the chosen configuration.

Here we discuss in more detail the LVIS MOT dynamics. Background gas atoms from the low-velocity tail of the thermal distribution are constantly cooled and pushed towards the confinement region of the trap. The opening in the output coupler defines a cylindrical region around the trap symmetry axis $z$, where an imbalance between the confining forces arises on both sides of the trap center. Therefore, an atom which ended up in the extraction column  will experience a net spontaneous force pointing towards the LVIS output coupler. The atoms from the resulting atom beam are continuously confined within the extraction column by the transverse laser beams (cooling and repumping). Those that diverge from the central column are recycled back into the trap. If the beam divergency, which is a measure for the atoms transverse velocity, is too large, the loading efficiency of the OM is compromised. Due to technical reasons we could not measure directly the atomic beam divergency in our experimental setup. Instead, we rely on observations and evaluations from previous studies carried out by Lu \cite{Wieman96} and Park \cite{Park99}. It was found that a pure geometrical factor can well describe the measured beam size and the divergence of the atomic beam scales as $\theta=d/x$, where $d$ is the diameter of the aperture in the LVIS output coupler and $x$ is the distance between the aperture and the LVIS trap center. In our case that would result in an atomic beam diameter of $15$\,mm at the OM trap center, where the OM cooling laser beams have a diameter of $42$\,mm ($1/e^2$).

\subsection{\label{sec:Mod}LVIS modification: Pump laser}

With the traditional LVIS setup \cite{Wieman96} and using transverse repumping, we achieved nearly 20 times more caesium atoms detected at the end of the fountain interrogation cycle in comparison with background gas OM loading. In both cases the loading time constant is about 1\,s. The efficiency of the LVIS--OM system is sensitive to the velocity of the beam atoms, because of the limited velocity capture range of the OM. However, the traditional LVIS system described so far does not provide enough flexibility to control the velocity of the atoms. The atoms in the extraction column will be accelerated (and heated) by the accelerating beam until the Doppler shift brings the atomic transition out of resonance \cite{Wang:03}. To stop the acceleration at a certain moment, it is expedient to shelve the atoms in a ``dark'' state  \cite{teo2002}, for which the $\ket{F=3}$ component of the caesium ground state is a convenient choice.

The only way for an atom in the extraction column to escape the cycling transition $\ket{F=4}$ $\rightarrow$ $\ket{F'=5}$ and to occur in the $\ket{F=3}$ state is by off-resonant excitation of the $\ket{F'=4}$ state. Since the frequency splitting between $\ket{F'=4}$ and $\ket{F'=5}$ is relatively large ($\sim 251$\,MHz), this process has low probability. Therefore, we bring into action an additional laser beam, called ``pump'' laser, which is intended to drive either the $\ket{F=4}$ $\rightarrow$ $\ket{F'=3}$ or $\ket{F=4}$ $\rightarrow$ $\ket{F'=4}$ transitions. While this laser grants a very efficient optical pumping to the $\ket{F=3}$ state, the presence of opposing repumping light along the $z$ axis would support continuous transfer of atoms back from the $\ket{F=3}$ to the $\ket{F=4}$ state during their flight to the OM zone. To avoid this scenario the repumping light is removed from the atomic beam path and only present in the vertically aligned LVIS MOT laser beams.

Fig.~\ref{LVIS} illustrates the present laser fields in our LVIS trap and their orientation. We have chosen the sign of the $B_{z}$ component of the magnetic field on both sides of the trap center ($z=0$) as shown in this figure. We define the quantization axis to be coincident with the $z$ axis and therefore the cooling laser light coming out of the fiber and having a wave vector pointing towards the LVIS output coupler must be $\sigma^{-}$ circularly polarized. The pump laser beam is spatially overlapped with the accelerating beam as they propagate in the same optical fiber, but it possesses opposite $\sigma^{+}$ circular polarization. Since beam atoms with positive displacement along $z$ ($B_z<0$) and outside the transverse laser region are shelved in the $\ket{F=4, m_{F}=-4}$ state by the on-axis $\sigma^{-}$ polarized cooling laser, only $\sigma^{+}$ polarized pump light can promote optical pumping to the $\ket{F=3}$ component according to the transition selection rules. As a result of the complementary laser pump field, the loaded number of atoms in the OM region significantly increases.

\begin{figure}
\includegraphics[width=0.48\textwidth]{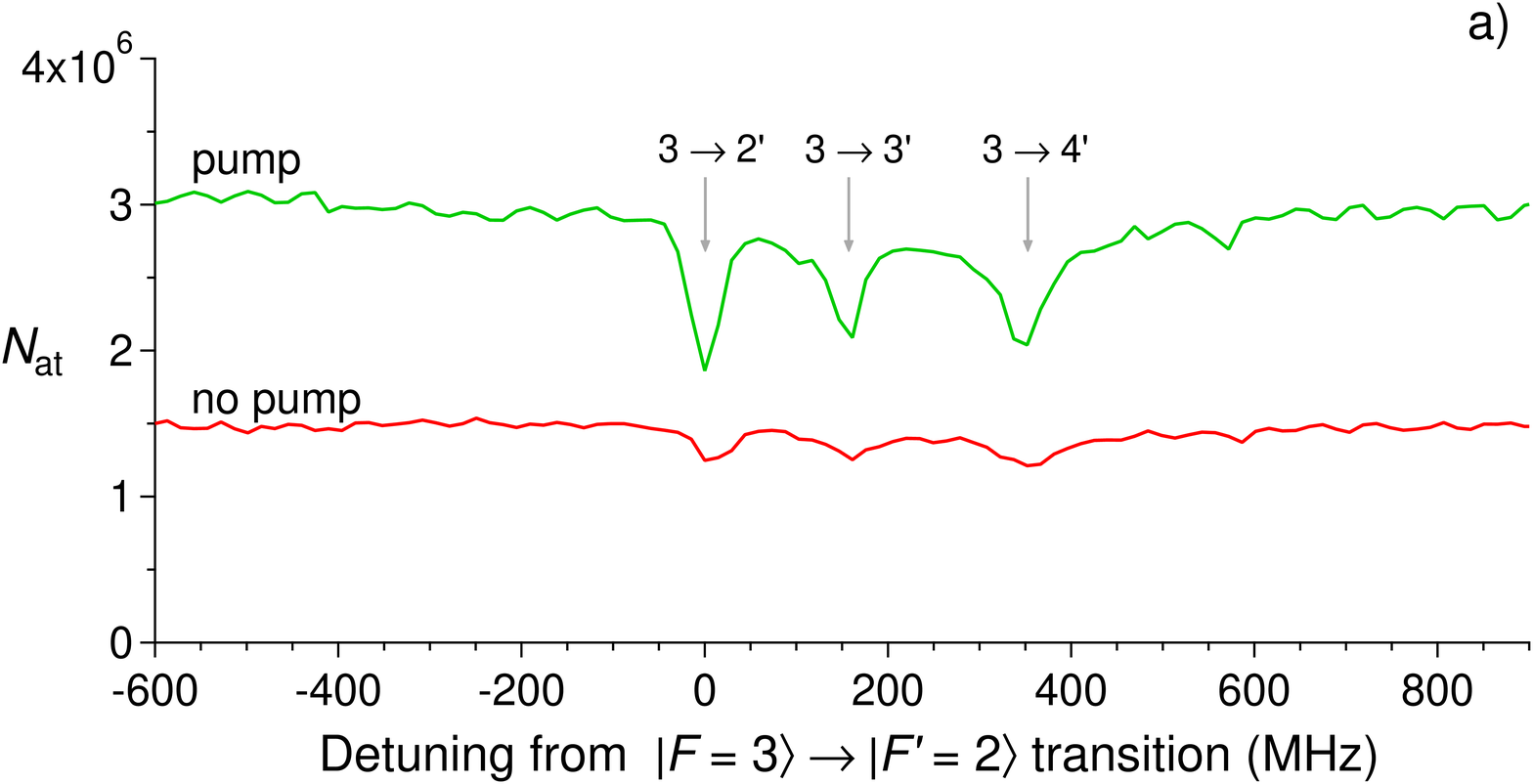}\vspace{1cm}
\includegraphics[width=0.48\textwidth]{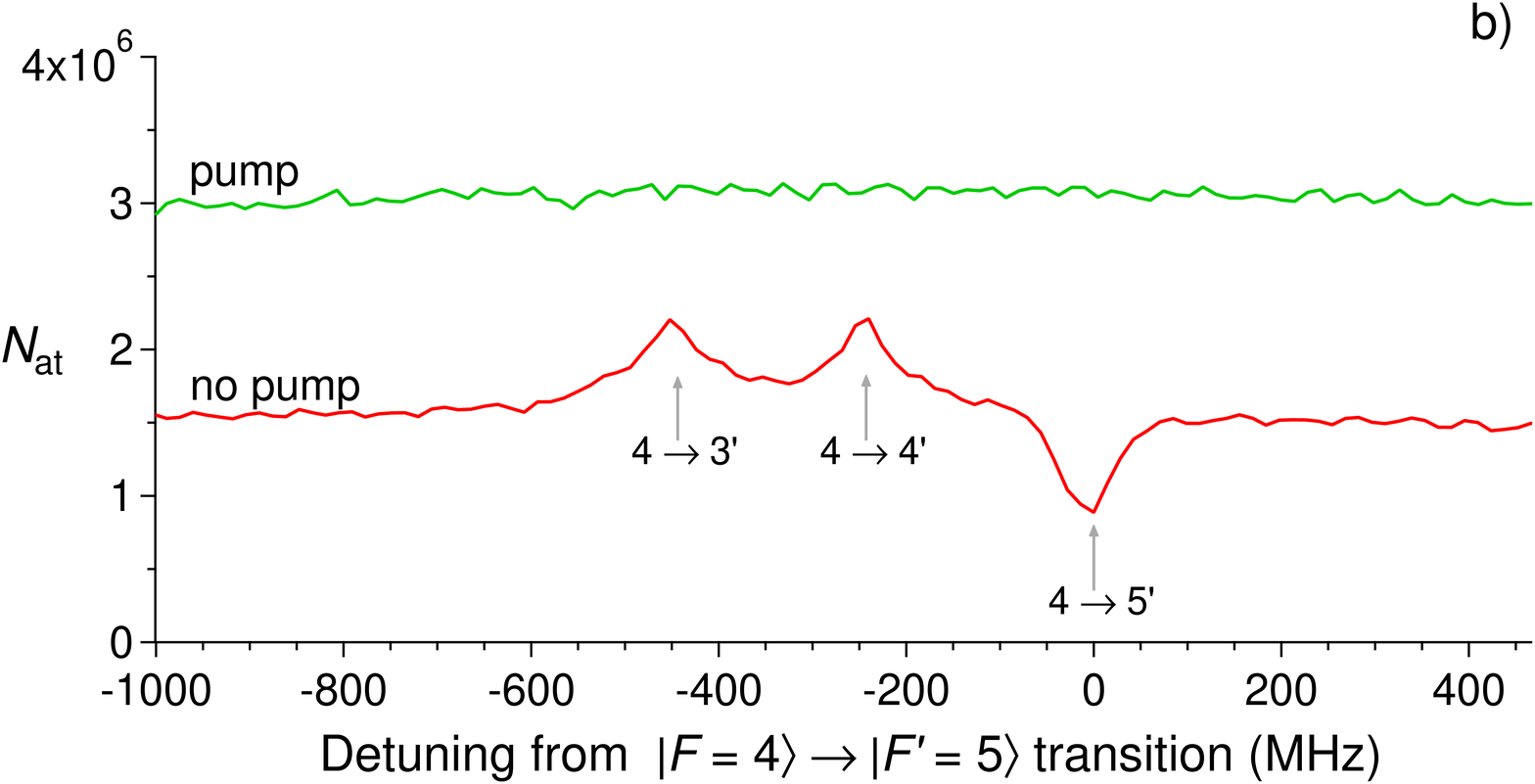}
\caption{\label{probe}The detected atom number $\textit{N}_{\mathrm{at}}$, resulting from fountain OM loading by means of the LVIS system, as a function of the probe laser frequency detuning from a) the $\ket{F=3}$ $\rightarrow$ $\ket{F'}$ and b) the $\ket{F=4}$ $\rightarrow$ $\ket{F'}$ transition. Results, with and without pump laser present, are shown by a green (top) and a red (bottom) trace, respectively. The pump laser is operated on the $\ket{F=3}$ $\rightarrow$ $\ket{F'=3}$ transition with an optical power of $0.01$\,mW. At exact resonance frequencies, quantum numbers $F'$ are indicated, while the crossover resonances in between are not denoted.}
\end{figure}

For verifying the internal atomic state of the atoms after they left the LVIS trap, a probe laser was applied perpendicular to the path of the atoms, 11\,cm away from the trap center. The probe laser was intended to transfer momentum to the atoms in a given $\ket{F}$ component in a direction transverse to that of propagation or to cause optical pumping. Both effects depend on the probe laser frequency detuning, polarization, intensity, and on the given static magnetic field. A rectangular aperture was used to produce a thin sheet of light from this laser with 2\,mm thickness and 10\,mm height, perpendicular to the atomic beam direction. In this way we ensured that the resulting probe light with 30\,$\mu$W optical power will interact with most of the passing atoms. In Fig.~\ref{probe} the detected atom number $N_{\mathrm{at}}$ is shown, when the frequency of the probe laser is scanned in the vicinity of both, the $\ket{F=3}$ $\rightarrow$ $\ket{F'=2}$ (Fig.~\ref{probe}a) and the $\ket{F=4}$ $\rightarrow$ $\ket{F'=5}$ (Fig.~\ref{probe}b) transitions.
In each sub-figure two modes of OM loading with the LVIS are shown. The top green and bottom red traces represent the situation with and without pump laser, respectively. In the following we discuss the results of Fig.~\ref{probe} only for a qualitative illustration of the most general features of the probe laser influence on the atom loading process, and do not attempt a quantitative explanation.

Without the pump laser (red traces in Fig.~\ref{probe}), there are atoms in both states $\ket{F=3}$ and $\ket{F=4}$, and their interaction with the probe laser tuned close to the $\ket{F}$ $\rightarrow$ $\ket{F'}$ resonances leads either to a decrease or an increase in the detected atom number $N_{\mathrm{at}}$. The atom number $N_{\mathrm{at}}$ is decreased, when either the probe laser pushes away beam atoms ($\ket{F=3}$ $\rightarrow$ $\ket{F'}$ (Fig.~\ref{probe}a) or $\ket{F=4}$ $\rightarrow$ $\ket{F'}$ (Fig.~\ref{probe}b) transitions), or pumps beam atoms from $\ket{F=3}$ to $\ket{F=4}$ ($\ket{F=3}$ $\rightarrow$ $\ket{F'=3,4}$ transitions, Fig.~\ref{probe}a)), so that they are subsequently detrimentally accelerated and heated by the cooling laser light. On the other hand the detected atom number increases, when the beam atoms are pumped by the probe laser from $\ket{F=4}$ to $\ket{F=3}$ ($\ket{F=4}$ $\rightarrow$ $\ket{F'=3,4}$ transitions, Fig.~\ref{probe}b), avoiding damaging acceleration and heating, and providing thus more efficient capturing in the fountain OM.

With the pump laser (green traces in Fig.~\ref{probe}), the atoms are transferred to the state $\ket{F=3}$, so that they do not interact with the probe laser tuned
around $\ket{F=4}$ $\rightarrow$ $\ket{F'=5}$ (Fig.~\ref{probe}b), while probe laser tuning close to the resonances $\ket{F=3}$ $\rightarrow$ $\ket{F'}$ (Fig.~\ref{probe}a) results in either pushing the atoms away by light scattering, or pumping them to $\ket{F=4}$ with subsequent detrimental acceleration by the cooling laser light.

\begin{figure}

    \includegraphics[width=0.48\textwidth]{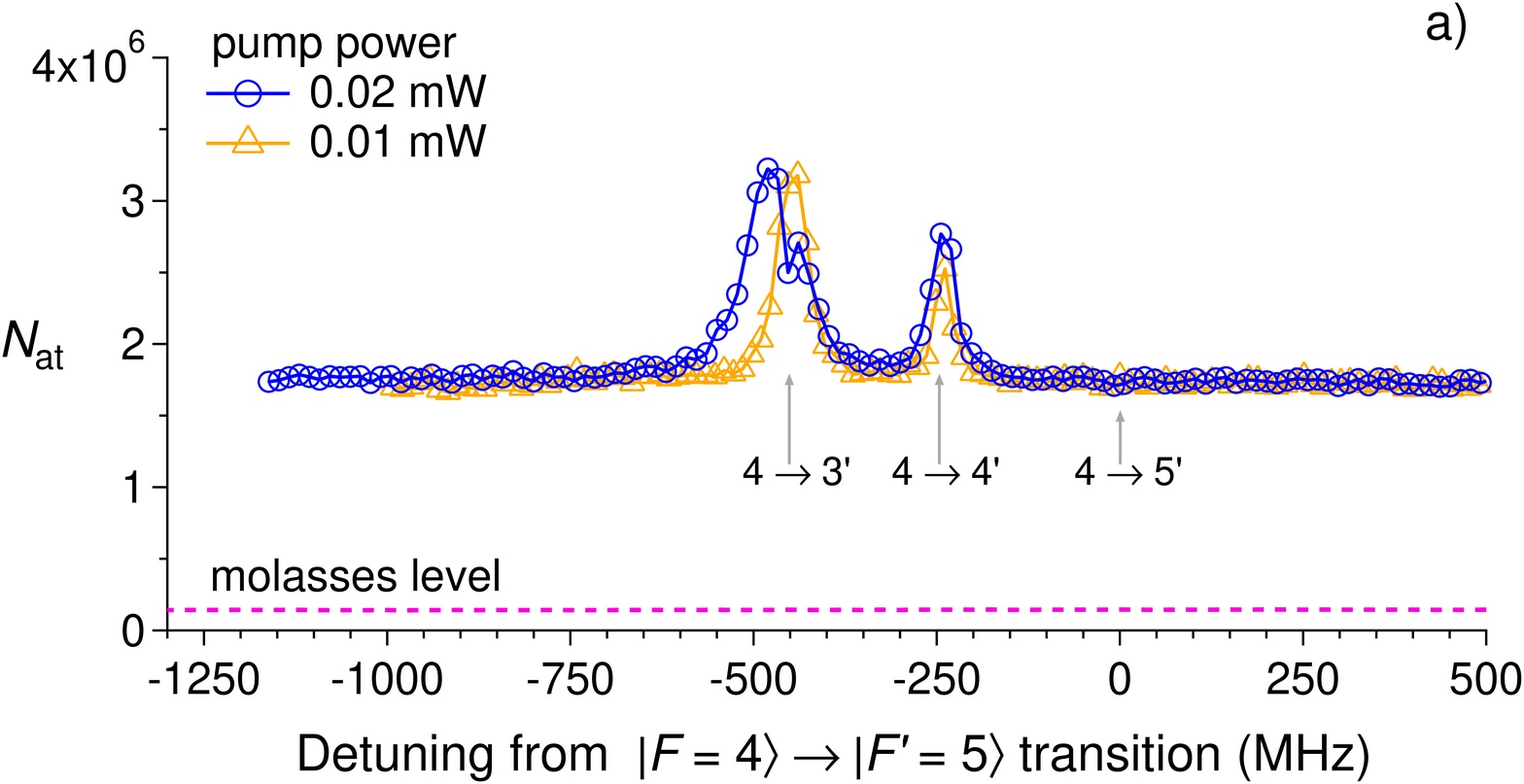}\vspace{1cm}
    \includegraphics[width=0.48\textwidth]{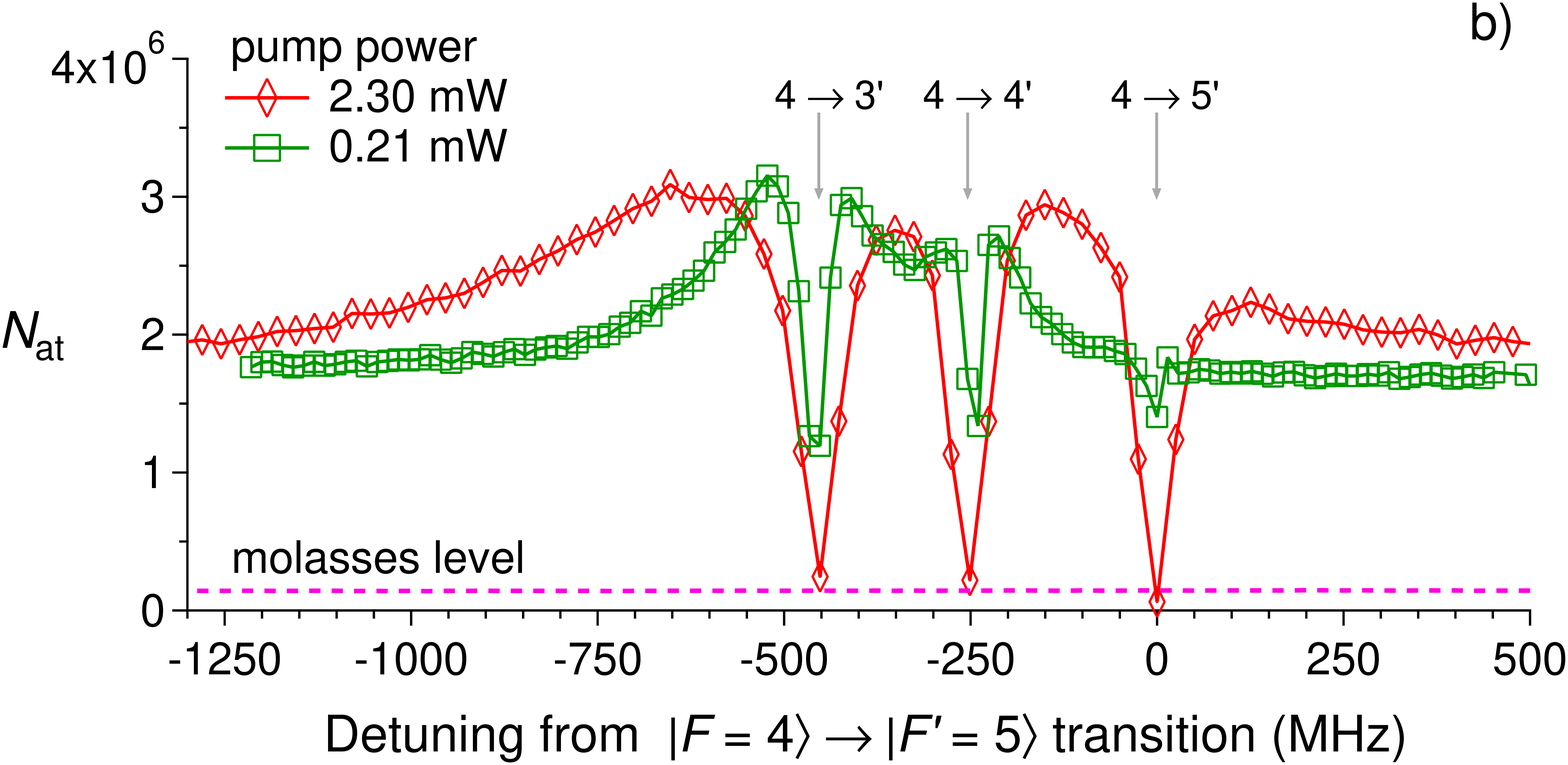}
    \caption{\label{holes}Plot of the detected atom number $N_{\mathrm{at}}$ as a function of the pump laser frequency at four different laser powers. For comparison, the dashed magenta trace represents $N_{\mathrm{at}}$ obtained in the standard mode of fountain operation (background gas OM loading).}

\end{figure}

In a next step we further investigate the effects of the pump laser properties. Figs.~\ref{holes}a) and~\ref{holes}b) illustrate the dependence of $N_{\mathrm{at}}$ on the pump laser frequency for several different values of its optical power. The laser frequency is continuously tuned through the $\ket{F=4}$ $\rightarrow$ $\ket{F'}$ resonances. For frequencies far-detuned from resonance, $N_{\mathrm{at}}$ stays stable at a level corresponding to the value of detected atoms without utilizing the pump laser. Once the frequency of the pump laser matches a transition, the process of optical pumping of atoms to the dark state starts to compete with the MOT loading rate. At low power (Fig.~\ref{holes}a) this laser efficiently transfers the atoms from the beam to the $\ket{F=3}$ component without significant heating so that most of them stay confined within the extraction column, and later on, travel unaffected towards the OM region. As a result, sharp peaks are observed at frequencies matching the $\ket{F=4}$ $\rightarrow$ $\ket{F'=3,4}$ transitions (orange triangle symbols in Fig.~\ref{holes}a). These peaks indicate the increased number of atoms which take part in the CSF2 measurement cycle. The difference in the transfer efficiency to the ``dark'' state noticed around the $\ket{F=4}$ $\rightarrow$ $\ket{F'=3}$ and $\ket{F=4}$ $\rightarrow$ $\ket{F'=4}$ resonances (Fig.~\ref{holes}a) can be explained by referring to the Clebsch-Gordan coefficients of the corresponding transitions \cite{Metcalf}. Because the pump laser beam has a very low optical power compared to the powers of the repumping and cooling laser beams (which are about three orders of magnitude higher), it does not significantly disturb the cooling process in the LVIS MOT.

When the power of the pump laser is increased, the cooling process becomes more disturbed, and a drop in $N_{\mathrm{at}}$ is observed. In the curves in Figs.~\ref{holes}a) (only $\ket{F=4}$ $\rightarrow$ $\ket{F'=3}$ at $\approx -450$\,MHz detuning, blue circles) and \ref{holes}b) (green squares and red rhomboids)
this appears as a formation of dips for frequencies close to the resonances. The dips widths broaden gradually when the power of the pump beam grows, since, even off-resonance, the disturbance of the LVIS MOT becomes more and more effective. At the same time, broad maxima of $N_{\mathrm{at}}$ develop (Figs.~\ref{holes}b)), more and more shifted to lower frequencies with respect to the $\ket{F=4}$ $\rightarrow$ $\ket{F'=3}$ and $\ket{F=4}$ $\rightarrow$ $\ket{F'=4}$ transitions. For sufficiently high pump optical power we even observe a complete loss of the LVIS MOT fluorescence and the number of detected atoms in CSF2 as well.
With the pump laser beam in a direction transverse to the LVIS symmetry axis there is no enhancement of $N_{\mathrm{at}}$. The observed dips in the spectrum close to the resonance frequencies were similar to the dips in Fig.~\ref{holes}b), caused by the same mechanisms of MOT disturbance. Such experiment also demonstrates that the optical pumping of the atoms to the $\ket{F=3}$ state mainly occurs after they leave the LVIS transverse laser beams. On the other hand, with the pump laser beam along the LVIS symmetry axis but with an opposite circular polarization $\sigma^{+}$ (the same as the cooling laser) also no increase of $N_{\mathrm{at}}$ is observed, as no optical pumping can take place according to transition selection rules. 

At the moderate pump optical powers (0.2\,mW to 2\,mW, see Fig.~\ref{holes}b) the maximum obtainable value of $N_{\mathrm{at}}$ is only $4\%$ lower compared to the maximum obtainable $N_{\mathrm{at}}$ at low pump optical powers (0.01\,mW to 0.02\,mW, see Fig.~\ref{holes}a). However, for moderate pump optical powers the loading process becomes less sensitive to the pump laser detuning. In our setup the peak value of the detected atom number reaches saturation at pump beam intensities of about $3.6$\,$\mu$W/cm$^{2}$ (50 $\mu$W power). For the frequency instability measurements (see Section~\ref{sec:Instability}) the power of the pump beam was 0.45\,mW with a red frequency detuning of 10\,$\Gamma$ from the $\ket{F=4}$ $\rightarrow$ $\ket{F'=3}$ transition.

We note that scanning the pump laser through the $\ket{F=3}$ $\rightarrow$ $\ket{F'}$ resonances shows no significant effect on $N_{\mathrm{at}}$ compared to the case of LVIS--OM loading without the pump laser - the pump laser effect on the atoms in the LVIS is similar to the effect of a repump laser along the $z$-axis. 


\begin{figure}

    \includegraphics[width=0.48\textwidth]{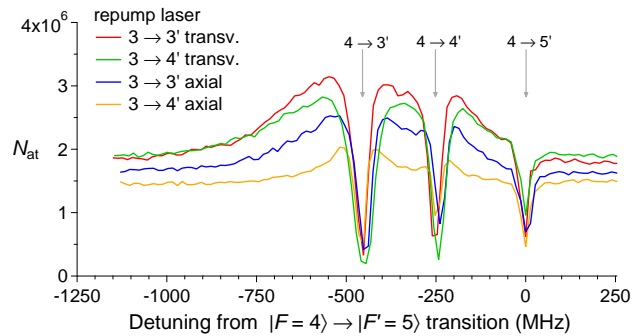}
      \caption{\label{repumper}Plot of the detected atom number $N_{\mathrm{at}}$ vs. the frequency of the pump laser for different LVIS repump laser configurations. Each configuration is characterized by the operating frequency of the repump laser (driving either the $\ket{F=3}$ $\rightarrow$ $\ket{F'=3}$ transition or the $\ket{F=3}$ $\rightarrow$ $\ket{F'=4}$ transition) and by the particular spatial orientation of the repumping light in the LVIS trap (either the repumping light is available along the $z$ axis (atomic beam axis) or it is applied only transversely to the $z$ axis). For all four cases the LVIS repump laser optical power was kept the same and the delivered power by the pump laser was $0.45$\,mW.}

\end{figure}

The properties of the LVIS repump laser also affect the loading process of the OM with the atomic beam. Fig.~\ref{repumper} illustrates the impact of both, operating frequency and orientation of the LVIS repump laser, on $N_{\mathrm{at}}$. Changing the repump laser tuning from the more efficient repumping transition $\ket{F=3}$ $\rightarrow$ $\ket{F'=4}$ to the less efficient repumping transition $\ket{F=3}$ $\rightarrow$ $\ket{F'=3}$ results in a higher number of cold atoms loaded into the OM of CSF2 (Fig.~\ref{repumper}). This finding is probably the result of two competing effects and depends on the particular repump laser intensities and geometry of our LVIS setup: While the utilisation of the $\ket{F=3}$ $\rightarrow$ $\ket{F'=4}$ repumping transition is beneficial for the LVIS MOT operation, stray light from the repump laser at this transition interacting with the $\ket{F=3}$ atoms in the beam is more efficient in pumping the atoms to the $\ket{F=4}$ state (resulting in the described detrimental acceleration effect) than stray light from the $\ket{F=3}$ $\rightarrow$ $\ket{F'=3}$ repumping transition. In our setup, with the given experimental parameters (pump laser power 0.45\,mW and a red frequency detuning of 10\,$\Gamma$ from the $\ket{F=4}$ $\rightarrow$ $\ket{F'=3}$ transition), the latter repumping transition is the better of the two choices, resulting in a factor of 40 increase of $N_{\mathrm{at}}$ compared to  operation of CSF2 with OM loading from background gas.

\subsection{\label{sec:Flux} Velocity and flux of the slow beam}

To characterize the source of cold atoms, we performed fluorescence and absorption measurements in the OM zone of the fountain. The first step in the atom velocity measurement was to allow the atomic beam to reach a steady state after the LVIS lasers were turned on. The probe laser was also turned on and tuned to the $\ket{F=3}$ $\rightarrow$ $\ket{F'=2}$ transition. Its intensity was increased to efficiently heat the cold beam atoms prepared in the $\ket{F=3}$ state by the pump laser, and to effectively prevent them from reaching the OM zone. After the atom beam flux reached a steady state ($0.5$\,s after the LVIS lasers were turned on), as a second step, the probe laser was turned off and the OM beams were simultaneously turned on for $1.1$\,s. The measured delay between the turning off of the probe beam (which allows the atoms to reach the OM zone) and the observed change in the OM loading rate (measured as sudden increase in the OM fluorescence) as the first cold atoms reach the OM zone was used to calculate a mean value for the atom velocity of $8.6$\,m/s.

We estimate the atom flux by performing a measurement of the cloud absorption in the OM zone. For this measurement, the OM and LVIS lasers were operated simultaneously for $1$\,s. At the end of this period all lasers were turned off and the probe beam was turned on, preventing additional arrival of cold atoms in the OM zone. After $10$\,ms, a weak absorption laser beam, resonant with the $\ket{F=4}$ $\rightarrow$ $\ket{F'=5}$ transition was introduced along the direction of one of the OM beams. It propagated in the OM fiber, and had the same profile and direction as the corresponding OM beam. The absorption beam power was $0.07$\,$\mu$W in a beam diameter of $42$\,mm ($1/e^2$). Its polarization was linear and orthogonal to that of the corresponding OM beam, and it was detected with a high-gain photodetector after passing the OM zone.

The absorption beam was kept on for $100$\,ms. During this measurement time, the absorption changed according to a loss rate consistent with losses due to gravity. From the measured relative absorption and the loading time of 1\,s, a mean atom flux from the enhanced LVIS system of $4\times10^{8}$/s is estimated.



\section{\label{sec:Instability} CSF2 operation with LVIS loading}

The frequency instability of a fountain frequency standard is expressed by the Allan deviation according to
\begin{equation}
\label{sy}
	\sigma_y\left(\tau\right)=\frac{1}{\pi}\frac{\Delta\nu}{\nu_0}\frac{1}{\text{SNR}}\sqrt{\frac{T_c}{\tau}},
\end{equation}
where $\Delta\nu$ is the full-width-at-half-maximum of the Ramsey fringe, $\nu_0=9\,192\,631\,770$\,Hz is the clock transition frequency, $\text{SNR}$ is the signal-to-noise ratio, $T_c$ is the cycle time, and $\tau$ the measurement time. In the case of quantum projection noise-limited operation, $\text{SNR}=\sqrt{N_{\mathrm{\mathrm{at}}}}$, with $N_{\mathrm{\mathrm{at}}}$ the total detected number of atoms in the $F=3$ and $F=4$ hyperfine components of the Cs ground state.

The $\text{SNR}$ was measured directly by operating the fountain CSF2 in a regime where the noise of the local oscillator does not contribute to the instability \cite{Santarelli1999}. The measured $\text{SNR}$ increases linearly as a function of $\sqrt{N_{\mathrm{\mathrm{at}}}}$ (measured in relative units) and reaches values larger than $1700$. The linear dependence between $\text{SNR}$ and $\sqrt{N_{\mathrm{\mathrm{at}}}}$ allows to calibrate $N_{\mathrm{\mathrm{at}}}$ in terms of the absolute number of detected atoms. The expected CSF2 instability, calculated from Eq.~\ref{sy} for $\Delta\nu=0.9$\,Hz, $T_c=2$\,s, $\text{SNR}=1680$ and $N_{\mathrm{at}}=2.9 \times 10^6$ is $\sigma_y\left(1\text{s}\right)=2.6 \times 10^{-14}$.

\begin{figure}

    \includegraphics[width=0.48\textwidth]{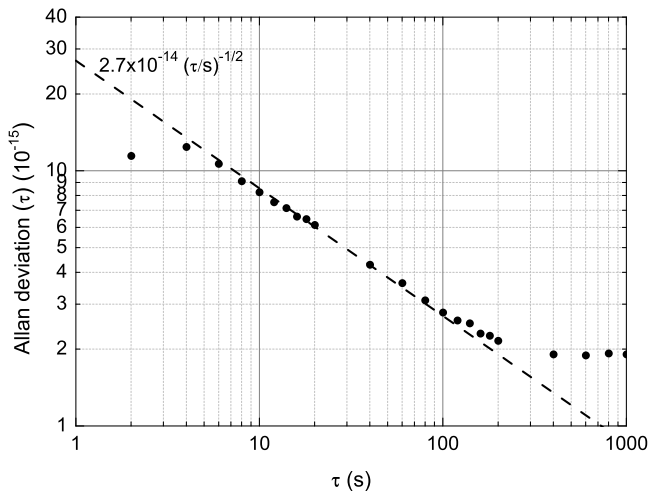}
      \caption{\label{instability} Allan deviation of the CSF2 frequency measured against the DRO referenced to an optical cavity (symbols). The fountain frequency instability of $\sigma_y\left(\tau\right) = 2.7 \times 10^{-14} \tau^{-1/2}$ is shown with a dashed line.}

\end{figure}

To experimentally confirm this value, the fountain was operated in a regime where the dominant contribution to its instability is the quantum projection noise. To reach this regime, an optically-stabilized 9.6\,GHz microwave signal is generated using a frequency comb as a transfer oscillator, and is used for the frequency synthesis of CSF2 \cite{Lipphardt2015}. The frequency comb is referenced to a laser locked to an ultra-stable optical cavity, transferring the stability of the laser to a 9.6\,GHz dielectric resonator oscillator (DRO). After removing the linear drift of the DRO frequency caused by the drift of the optical cavity, the measured Allan deviation $\sigma_y\left(\tau\right)$ shows a $\tau^{-1/2}$ dependence for measurement times up to 100\,s (Fig.~\ref{instability}).

The measured fountain instability $\sigma_y\left(1\text{s}\right)=2.7 \times 10^{-14}$ is in good agreement with the value of $2.6 \times 10^{-14}$ inferred from the $\text{SNR}$ measurements. It is also close to the best measured instability of $\sigma_y\left(1\text{s}\right)=1.6 \times 10^{-14}$ in a primary fountain clock, where the atoms were loaded from a decelerated Cs atomic beam \cite{Vian2005}.


\section{\label{sec:Conclusions} Conclusions}

Slow atomic beam loading of the PTB fountain CSF2 is demonstrated. The source of cold atoms is a modified low-velocity intense source setup which includes an additional pump laser. The pump laser reduces the loss of cold atoms during their flight between the LVIS apparatus and the OM loading zone by pumping them into a dark state in which they are not subject to continued acceleration and heating due to light scattering. Additionally, the atoms are loaded in a volume in the proximity of the fountain axis by using a repump laser beam only propagating along the fountain axis, which further increases the number of detected atoms and potentially reduces the contribution of the distributed cavity phase to the uncertainty of CSF2 \cite{Weyers2012b}. The achieved detected atom number is a factor of 40 above the one used in normal operation of CSF2. The LVIS atom flux and velocity are characterized through fluorescence and absorption measurements. With the LVIS in operation and using an optically-stabilized microwave signal, the fountain frequency instability is quantum projection noise-limited and has a value of $\sigma_y\left(1\text{s}\right)=2.7 \times 10^{-14}$. With such instability, the statistical uncertainty of the fountain reaches the level of its present systematic uncertainty in $10^4$\,s.

\begin{acknowledgments}
    The authors thank B. Lipphardt, N. Huntemann and M. Okhapkin for valuable discussions, and D. Griebsch and N. Nemitz for designing the initial LVIS setup. This work was supported by the European Metrology Research Programme (EMRP) in projects SIB04 and SIB55. The EMRP is jointly funded by the EMRP participating countries within EURAMET and the European Union.
\end{acknowledgments}

\bibliography{LVIS_Bibliography}

\end{document}